\documentclass[12pt]{article}
\usepackage{times}

\def\beq{\begin{equation}}
\def\eeq{\end{equation}}
\def\bea{\begin{eqnarray}}
\def\eea{\end{eqnarray}}

\def\la{\langle}
\def\ra{\rangle}

\def\nn{\nonumber}
\def\Eq#1{Eq.~(\ref{#1})}

\def\x#1{\langle#1\rangle}

\def\xx#1{\langle\!\langle#1\rangle\!\rangle}
\newcommand{\ta}[1]{#1\hspace{-.42em}/\hspace{-.07em}}

\begin{document} 

\begin{titlepage}

\renewcommand{\thefootnote}{\fnsymbol{footnote}}
\begin{flushright}
    IFIC/06-03   \\ hep-th/0602043
\end{flushright}
\par \vspace{10mm}

\begin{center}

{\Large \bf Compact multigluonic scattering amplitudes 
with heavy scalars and fermions}

\vspace{8mm}

{\bf Paola Ferrario~$^{(a)}$\footnote{E-mail: paola.ferrario@ific.uv.es}},
{\bf Germ\'an Rodrigo~$^{(a)}$\footnote{E-mail: german.rodrigo@ific.uv.es}}
and
{\bf Pere Talavera~$^{(b)}$\footnote{E-mail: pere.talavera@upc.edu}}

\vspace{5mm}
${}^{(a)}$ Instituto de F\'{\i}sica Corpuscular, 
CSIC-Universitat de Val\`encia, \\
Apartado de Correos 22085, 
E-46071 Valencia, Spain. \\

\vspace{5mm}

${}^{(b)}$ Departament de F\'isica y Enginyeria Nuclear,  
Universitat Polit\`ecnica de Catalunya, Jordi Girona 1-3, 
E-08034 Barcelona, Spain. \\

\vspace{5mm}

\end{center}

\par \vspace{2mm}
\begin{center} {\large \bf Abstract} \end{center}
\begin{quote}
\pretolerance 10000
Combining the Berends-Giele and on-shell recursion relations 
we obtain an extremely compact expression for the 
scattering amplitude of a complex scalar-antiscalar 
pair and an arbitrary number of positive helicity 
gluons. This is one of the basic building blocks for 
constructing other helicity configurations from recursion 
relations. We also show explicity that  the all 
positive helicity gluons amplitude for heavy fermions 
is proportional to the scalar one, confirming in this way
the recently advocated SUSY-like Ward identities relating 
both amplitudes.
\end{quote}

\vspace*{\fill}
\begin{flushleft}
     IFIC/06-03 \\ 
     February 6, 2006
\end{flushleft}
\end{titlepage}

\setcounter{footnote}{1}
\renewcommand{\thefootnote}{\fnsymbol{footnote}}


\section{Motivation}

To achieve a  successful physics program at LHC there must be a good 
control over all the possible expected backgrounds. 
These, among other processes, require the evaluation of 
multipartonic scattering amplitudes at higher orders
in the perturbative expansion.  Without this information the 
identification of any signal of new physics is only partial. 
Despite their relevance, Yang-Mills scattering amplitudes are very 
poorly known, mainly because the number of Feynman diagrams 
increases exponentially with the number of the external fields 
involved in the processes. This is one of the main reason for 
elaborating  other techniques to obtain the amplitudes. 
Between them one of the most successful is the use of 
recursion relations within the helicity amplitude formalism. 
The helicity amplitude formalism~\cite{Jacob:1959at,Bjorken:1966kh,Mangano:1990by}
has been proven to be an elegant and efficient tool
to calculate multipartonic scattering amplitudes.
Recursion relations  extensively used in the 
literature at tree~\cite{Berends:1987me,Dixon:1996wi}
and one-loop level~\cite{Bern:1994zx,Bern:1994cg} to 
calculate multipartonic scattering amplitudes.

Based on old insights, \cite{Nair:1988bq},  in Ref.~\cite{Witten:2003nn} 
Witten presents the idea of a weak-weak duality between supersymmetric 
${\cal N} =4$ Yang-Mills and topological B string theories in twistor space.
Inspired by, but independent of these findings, a new method for the 
evaluation of scattering amplitudes 
in gauge theories has been proposed~\cite{Cachazo:2004kj}, 
the so called CSW. It is based on the recursive use of off-shell 
Maximal Helicity Violating amplitudes (MHV)~\cite{Parke:1986gb} as basic
vertices for new amplitudes. 
Recent works have accomplished interesting progress since the original 
formulation, and the method has been refined by 
introducing more efficient recursion relations~\cite{Britto:2004ap,Britto:2005fq}, 
the so called BCFW, 
and extending this approach to the one-loop level~\cite{Bern:2005cq,Bern:2005hs}. 

Extending the BCFW formalism to massive particles, 
on-shell recursion relations at tree-level have 
been introduced in Ref.~\cite{Badger:2005zh} for massive scalars, 
and in Ref.~\cite{Badger:2005jv} for vector boson and fermions. 
Scattering amplitudes with heavy scalars and up to four gluons of 
positive helicity were first derived in Ref.~\cite{Bern:1996ja}.
In Ref.~\cite{Badger:2005zh} all the helicity configurations 
with up to four gluons have been computed by using the 
on-shell recursion relations. 
These results have been extended to amplitudes with an arbitrary number 
of gluons of identical helicity or one gluon of opposite helicity 
in Ref.~\cite{Forde:2005ue}. The approach of 
Forde and Kosower~\cite{Forde:2005ue} is based on a basic ansatz for 
the all positive helicity amplitude which is shown to 
fullfil the BGKS~\cite{Badger:2005zh} recursion relations, 
and which is used to construct the rest of the helicity configurations.
Using off-shell recursion relations~\cite{Berends:1987me},
multigluonic scattering amplitudes with heavy fermions
and an arbitrary number of gluons of positive helicity 
have been calculated in Ref.~\cite{Rodrigo:2005eu}. 

We are concern in this note with two kind of 
multigluonic scattering amplitudes, and more in concrete with their relation: 
the first involves heavy fermions and are 
interesting by its own, due the expected rich phenomenology driven 
by the heavy quarks at LHC. The second of the amplitudes, with complex colored massive scalars, 
are of use in the unitarity method for computing massless loop 
amplitudes in nonsupersymmetric gauge theories~\cite{Bern:2005cq}. 
In a recent paper~\cite{Schwinn:2006ca}  it has been demonstrated 
that a SUSY-like model Ward identities relate both amplitudes with 
heavy scalars and fermions. The apparent quite different 
structure of the results presented in Refs.~\cite{Forde:2005ue}
and~\cite{Rodrigo:2005eu} makes however quite difficult to test 
explicitly that relationship, apart for amplitudes with a few 
gluons due to their simplicity. 

Is our aim to show explicitly inside QCD, that for a given helicity
configuration the multigluonic massive heavy quark and the massive 
heavy scalar amplitudes are related by a simple overall kinematical factor. 
For this we construct in Sec. (2) the off-shell massive scalar amplitude. In Sec. (3) we review the equivalent fermionic amplitude and present the relation with the scalar case. Finally Sec. (4) contains our summary.  Some notations and definitions issues are gathered in an Appendix.

\section{Scalar amplitudes}

The colour ordered off-shell current of an on-shell complex scalar of 
four-momentum $p_1$ and $(n-2)$-gluons of four-momenta $p_2$ to $p_{n-1}$ 
and positive helicity is given in terms of the off-shell scalar current 
with less gluons, and the off-shell gluonic current $J^\mu$
of the rest of the gluons: 
\beq
S(1_s;2^+,\ldots,n-1^+) = - \frac{\sqrt{2}}{y_{1,n-1}} \, \sum_{k=1}^{n-2}
S(1_s;2^+,\ldots,k^+) \, p_{1,k} \cdot J(k+1^+,\ldots,n-1^+)~,
\label{scalarrecursion}
\eeq
where $p_{1,k}=p_1+p_2+\ldots +p_k$ and $S(1_s) = 1$. 
We also define $y_{1,k}=p_{1,k}^2-m^2$.
For all gluons of positive helicity the gluonic current 
has the form~\cite{Berends:1987me}:
\beq
J^\mu(i^+,\ldots,j^+) = 
\frac{\x{\xi|\gamma^\mu \ta{p}_{i, j}|\xi}}
{\sqrt{2}\x{\xi i}\xx{i,j} \x{j\xi}}~,
\eeq
where 
\beq
\xx{i,j} = \x{i(i+1)}\x{(i+1)(i+2)}\cdots \x{(j-1)j}~,
\eeq
with $\xx{i,i} = 1$~.
The null vector $\xi$ is the reference gauge vector which is assumed
to be the same for all the gluons. 
Then from \Eq{scalarrecursion}, we get
\beq
S(1_s;2^+,\ldots,n-1^+) = - \frac{1}{y_{1,n-1} \, \x{(n-1)\xi}} \, \sum_{k=1}^{n-2}
S(1_s;2^+,\ldots,k^+) \, \frac{\x{\xi|\ta{p}_{1,k}\ta{p}_{k+1,n-1}|\xi}}
{\x{\xi (k+1)} \xx{k+1,n-1}}~.
\label{recursion2}
\eeq
To obtain the recursion relation in \Eq{scalarrecursion}
we apply the Berends-Giele rules~\cite{Berends:1987me}
and consider the $\phi g \phi^\dagger$ vertex 
\beq
V(p_1,k^\mu,p_2) = \frac{1}{\sqrt{2}}(p_2-p_1)^\mu~,
\eeq
where $p_1$, $k$ and $p_2$ are the four-momenta of the scalar, 
the gluon and the antiscalar respectively, and the $\sqrt{2}$
comes from the normalization conventions used in colour 
ordered Feynman rules.  
Four-point vertices do not contribute to the current with 
all the gluons of the same helicity, since 
\beq
J(i^+,\ldots,j^+)\cdot J(k^+,\ldots,l^+) = 0~.
\eeq

Let's anticipate our result for the scalar current with 
an arbitrary number of gluons:
\bea
S(1_s;2^+,\ldots,n-1^+) &=& 
\frac{\x{(n-2)\xi}}{\x{(n-2)(n-1)} \, \x{(n-1)\xi}} 
\, S(1_s;2^+, \ldots, n-2^+) \nn \\ 
&+& \frac{i}{y_{1,n-1}} \, A_n(1_s;2^+,\ldots,n-1^+;n_s)~,
\label{offshell}
\eea
where
\bea
&& \!\!\!\!\!\!\!
A_n(1_s;2^+,\ldots,n-1^+;n_s) = i \, 
\frac{m^2}{y_{12} \, y_{1,3} \, \xx{2,n-1} }
\Bigg\{ [2|\ta{p}_1 \ta{p}_{23}|n-1] \nn \\
&& + \sum_{j=1}^{n-5} \, [2|\ta{p}_1 \ta{p}_{23}|w_1] \,
\frac{\la w_1| \ta{p}_{1,w_1-1} |w_2]}{-y_{1,w_1}} \cdots 
\frac{\la w_j| \ta{p}_{1,w_j-1} |n-1]}{-y_{1,w_j}} \Bigg\}~,
\label{onshell}
\eea
with $w_1<w_2< \dots < w_j$ and $w_k \in [4,\ldots,n-2]$,
is the corresponding on-shell amplitude which 
is obtained from the off-shell current
by removing the propagator of the off-shell
antiscalar, and imposing momentum conservation: $y_{1,n-1}=0$.
The well known one-, two-, and three-gluon on-shell scattering 
amplitudes are 
\beq
A_3(1_s;2^+;3_s) = i \, \frac{\la \xi|\ta{p}_1|2]}{\x{\xi2}}~,
\eeq
\beq
A_4(1_s;2^+,3^+;4_s) = i \, \frac{m^2[23]}{y_{12} \, \x{23}}~,
\eeq
\beq
A_5(1_s;2^+,3^+,4^+;5_s) = i \, \frac{m^2[2|\ta{p}_1\ta{p}_{23}|4]}
{y_{12} \, y_{1,3} \xx{2,4}}~.
\eeq

To obtain these results we have performed the following transformation 
in the first term of \Eq{recursion2}: 
\beq
\x{\xi| \ta{p}_1 \ta{p}_{2,n-1}| \xi} = \frac{1}{y_{12}}
\left( m^2 \x{\xi 2} [2 | \ta{p}_{3,n-1}|\xi \ra
+ \la \xi| \ta{p}_1 | 2] \x{2 | \ta{p}_1 \ta{p}_{2,n-1}|\xi}
\right)~,
\eeq
together with 
\beq
[2 | \ta{p}_{3n-1}|\xi \ra = \frac{1}{y_{1,3}}
\left( [2 | \ta{p}_1 \ta{p}_{23} \ta{p}_{4,n-1}|\xi \ra
- [32] \x{3| y_{12}+ \ta{p}_{12} \ta{p}_{3,n-1} | \xi}
\right)~.
\eeq
Because of the Schouten identity the rest of the terms 
can be written as 
\bea
\x{\xi|\ta{p}_{1,k}\ta{p}_{k+1,n-1}|\xi} &=& \frac{1}{\x{(n-2)(n-1)}} 
\bigg( \x{\xi(k+1)}
\x{k|y_{1,k-1}+\ta{p}_{1,k-1}\ta{p}_{k,n-1}|\xi} \nn \\
&-& \x{\xi k} \x{k+1|y_{1,k}+\ta{p}_{1,k}\ta{p}_{k+1,n-1}|\xi}\bigg)~,
\label{schouten}
\eea
with $y_{1,1}=0$, and 
\beq  
\x{n-1|y_{1,n-2}+\ta{p}_{1,n-2}\ta{p}_{n-1}|\xi} = - y_{1,n-1} \x{\xi (n-1)}~.
\eeq
The latter generates the first term in \Eq{offshell} that cancels in 
the on-shell amplitude. Finally, we use
\beq
\x{k | \ta{p}_{1,k-1} \ta{p}_{k,n-1} | \xi }
= \sum_{j=k}^{n-1} \, \la k | \ta{p}_{1,k-1} | j ] \x{j\xi}~, 
\label{nogauge}
\eeq
to remove the gauge dependence of the on-shell amplitude.

The number of terms in \Eq{onshell} grows as $2^{n-5}$, but 
contrary to the Forde-Kosower's ansatz for that amplitude~\cite{Forde:2005ue}  
our expression do not contain different powers of the mass,
being always proportional to $m^2$. This fact makes
easier the validation of our expression through the on-shell BGKS 
recursion relations, and allows us to relate the scalar amplitude 
with the fermionic one in a straightforward way. 
As in Ref.~\cite{Forde:2005ue} we perform a shift in the four-momenta 
of the (2,3) gluons:
\bea
&& \hat{p}_2^\mu = p_2^\mu + \frac{z}{2} [2|\gamma^\mu|3\ra~,\nn \\
&& \hat{p}_3^\mu = p_3^\mu - \frac{z}{2} [2|\gamma^\mu|3\ra~.
\eea
That shift corresponds to the following shift of the spinors
\bea
& |\hat{2}\ra = |2\ra + z |3\ra~, & |\hat{2}] = |2]~, \nn \\
& |\hat{3}] = |3] - z |2]~, & |\hat{3}\ra = |3 \ra~.
\eea
The only term that contributes to the recursion 
relation is the one where the scalar and the first gluon
are factorized in the left side:
\beq
A_n(1_s^+;2^+, \ldots, n-1^+;n_{s}) = 
A_3(1_s^+;\hat{2}^+; -\hat{p}_{12s}) 
\, \frac{i}{y_{12}} \,  
A_{n-1}(\hat{p}_{12s};\hat{3}^+, \ldots, n-1^+;n_{s})~.  
\label{bgks}
\eeq
Thus, we have 
\bea
&& \!\!\!\!\!\!\!
A_n(1_s;2^+,\ldots,n-1^+;n_s) = i \, 
\frac{m^2 \, [2|\ta{p}_1 \, \ta{\hat{p}}_3 \, \ta{\hat{p}}_{12} \, \ta{\hat{p}}_{34}}
{y_{12} \, y_{1,3} \, y_{1,4} \, \xx{2,n-1} }
\Bigg\{ |n-1] \nn \\
&& + \sum_{j=1}^{n-6} \, |w_1] \,
\frac{\la w_1| \ta{p}_{1,w_1-1} |w_2]}{-y_{1,w_1}} \cdots 
\frac{\la w_j| \ta{p}_{1,w_j-1} |n-1]}{-y_{1,w_j}} \Bigg\}~,
\label{scalarbcfw}
\eea
with the gauge choice $\xi=\hat{3}$ in the left amplitude, 
and $w_k \in [5,\ldots,n-2]$. 
For the channel under consideration $z=- y_{12}/[2|\ta{p}_1|3\ra$. 
Using this value for the shifted four-momenta 
the following relation holds after some algebra 
\beq
[2|\ta{p}_1 \, \ta{\hat{p}}_3 \, \ta{\hat{p}}_{12} \, \ta{\hat{p}}_{34} =
[2| \ta{p}_1 \ta{p}_{23} \, (y_{1,4} - \ta{p}_4 \, \ta{p}_{1,3})~.
\label{resimple}
\eeq
Then, with the help of \Eq{resimple} it becomes 
almost trivial to demonstrate that \Eq{onshell} fullfils 
the on-shell recursion relation in \Eq{bgks}. The first term 
in the rhs of \Eq{resimple} generates all the terms that do not contain the 
$1/y_{1,4}$ propagator, the second term instead initiates the 
spinorial chains for which $w_1=4$. This fact also explains why the 
number of terms contributing to the amplitude doubles each 
time that we add one extra gluon. 
On the other hand, it is worth to notice that we can bring 
the lhs of the rhs of \Eq{resimple} into the form 
\beq
[2|\ta{p}_1 \ta{p}_{23} = [2| (y_{1,3}- \ta{p}_3 \ta{p}_{12})~.
\eeq
This suggest that we can either extend the sum in \Eq{onshell}
down to $w_k=3$, or even better, we can regroup all the terms in 
the sum into a single one by going upwards. 
Our final result for the amplitude with 
all gluons of positive helicity 
becomes in this way extremely compact:
\bea
&& \!\!\!\!\!\!\!
A_n(1_s;2^+,\ldots,n-1^+;n_s) = i \, m^2 \, 
\frac{[2| \prod_{k=3}^{n-2} (y_{1,k}-\ta{p}_k \ta{p}_{1,k-1}) |n-1]}
{y_{12} \, y_{1,3} \cdots y_{1,n-2} \, \xx{2,n-1} }~.
\label{recompact}
\eea

\section{Fermionic amplitudes}

The all positive helicity gluon amplitudes with a heavy 
fermion-antifermion pair have been calculated 
in Ref.~\cite{Rodrigo:2005eu}. We have worked out 
further these expressions with the help of \Eq{schouten} and \Eq{nogauge}
in order to obtain a more compact formulae to compare 
with the heavy scalar amplitude. With our 
spinor choice the on-shell helicity conserving amplitude 
vanishes, and for the helicity flip amplitude we
find in a straightforward way the following relationship
to the scalar amplitude
\bea
A_n(1_q^+;2^+, \ldots, n-1^+;n_{\bar{q}}^+) &=& 
\frac{m}{\beta_+ \x{1n}} \, A_n(1_s;2^+, \ldots, n-1^+;n_s)~,
\eea
with $\beta_+$ as given in the Appendix.
This represents an explicit and independent confirmation 
of the SUSY-like Ward identities found recently 
in Ref.~\cite{Schwinn:2006ca}, that relate several 
multigluonic amplitudes of heavy scalars and fermions. 
Since we have obtained a very compact expression for
the scalar amplitude, the same simple result holds 
for the case of heavy fermions. 

\section{Summary}

Combining off-shell and on-shell recursion relations 
we have obtained an extremely compact expression for 
the scattering amplitude of a colored scalar-antiscalar
pair and an arbitrary number of gluons of positive 
helicity at tree-level. We think that  Eq.~(\ref{recompact}) is the 
most reduced expression one can obtain for such process.  
This result is the main input to obtain
other helicity configurations from recursion relations. 
Due to its simplicity, we expect also that these other
amplitudes can be calculated more efficiently and 
will be written in a more compact way 
than previously published. SUSY-like Ward identities 
might also help to extend these simple results to 
amplitudes with heavy fermions, or viceversa. In particular, 
we have tested explicity the validity of these identities 
relating scalar and fermionic amplitudes with an 
arbitrary number of positive helicity gluons. 
Eventhough these kind of relations are so far valid just at tree level.

\section*{Acknowledgements}

This work has been partially supported by Ministerio de Educaci\'on y Ciencia (MEC) 
under grants FPA2004-00996  and FPA 2004-04582-C02-01,
Acciones Integradas DAAD-MEC (contract HA03-164), Generalitat Valenciana (GV05-015), by the European 
Commission RTN Program (MRTN-CT-2004-005104, MRTN-CT-2004-503369), and by the Generalitat de Catalunya (CIRIT GC 2001SGR-00065).

\appendix

\section{Spinors and heavy four-momenta}

We follow the conventions of Ref.~\cite{Rodrigo:2005eu},
and denote by $p_1^\mu$ and $p_n^\mu$, with $p_1^2=p_n^2=m^2$, 
the four-momenta of the heavy particles. 
In terms of two light-like vectors ($\bar{p}_1^2=\bar{p}_n^2=0$)
these four-momenta can be written as 
\bea
p_1^\mu &=& \beta_+ \, \bar{p}_1^\mu + 
\beta_- \, \bar{p}_n^\mu~, \nn \\
p_n^\mu &=& \beta_- \, \bar{p}_1^\mu + 
\beta_+ \, \bar{p}_2^\mu~, 
\eea
where $\beta_\pm=(1\pm \beta)/2$
with $\beta=\sqrt{1-4m^2/s_{1n}}$ the velocity of the heavy particles, 
and $s_{1n} = (p_1+p_n)^2$. Among other advantages, this  
transformation preserves momentum conservation such that 
$p_1+p_n=\bar{p}_1+\bar{p}_n$.
Furthermore, in the massless limit we have: $p_1\to \bar{p}_1$ 
and $p_2\to \bar{p}_2$. 

If the heavy particles are fermions, we use the following 
choice  of spinors 
\beq
\bar{u}_\pm(p_1,m) = \frac{\beta_+^{-1/2}}{\x{n^\mp 1^\pm}} 
\langle n^\mp | \, (\ta{p}_1+m)~, \qquad
v_\pm(p_n,m) = \frac{\beta_+^{-1/2}}{\x{n^\mp 1^\pm}}
(\ta{p}_n-m) \, |1^\pm \rangle~,
\eeq
where $| i^\pm \rangle = | \bar{p}_i^\pm \rangle$ are the 
Weyl spinors of the light-like vectors.


\begin{thebibliography}{90}


\bibitem{Jacob:1959at}
  M.~Jacob and G.~C.~Wick,
  Annals Phys.\  {\bf 7} (1959) 404
  [Annals Phys.\  {\bf 281} (2000) 774].

\bibitem{Bjorken:1966kh}
  J.~D.~Bjorken and M.~C.~Chen,
  Phys.\ Rev.\  {\bf 154} (1966) 1335.


\bibitem{Mangano:1990by}
  M.~L.~Mangano and S.~J.~Parke,
  Phys.\ Rept.\  {\bf 200} (1991) 301.


\bibitem{Berends:1987me}
  F.~A.~Berends and W.~T.~Giele,
  Nucl.\ Phys.\ B {\bf 306} (1988) 759.

\bibitem{Dixon:1996wi}
  L.~J.~Dixon,
  arXiv:hep-ph/9601359.

\bibitem{Bern:1994zx}
  Z.~Bern, L.~J.~Dixon, D.~C.~Dunbar and D.~A.~Kosower,
  Nucl.\ Phys.\ B {\bf 425} (1994) 217 
  [arXiv:hep-ph/9403226].

\bibitem{Bern:1994cg}
  Z.~Bern, L.~J.~Dixon, D.~C.~Dunbar and D.~A.~Kosower,
  Nucl.\ Phys.\ B {\bf 435} (1995) 59 
  [arXiv:hep-ph/9409265].



\bibitem{Nair:1988bq}
  V.~P.~Nair,
  Phys.\ Lett.\ B {\bf 214} (1988) 215.

\bibitem{Witten:2003nn}
  E.~Witten,
  Commun.\ Math.\ Phys.\  {\bf 252} (2004) 189
  [arXiv:hep-th/0312171].

\bibitem{Cachazo:2004kj}
  F.~Cachazo, P.~Svrcek and E.~Witten,
  JHEP {\bf 0409} (2004) 006
  [arXiv:hep-th/0403047].


\bibitem{Parke:1986gb}
  S.~J.~Parke and T.~R.~Taylor,
  Phys.\ Rev.\ Lett.\  {\bf 56} (1986) 2459.


\bibitem{Britto:2004ap}
  R.~Britto, F.~Cachazo and B.~Feng,
  Nucl.\ Phys.\ B {\bf 715} (2005) 499
  [arXiv:hep-th/0412308].

\bibitem{Britto:2005fq}
  R.~Britto, F.~Cachazo, B.~Feng and E.~Witten,
  Phys.\ Rev.\ Lett.\  {\bf 94} (2005) 181602
  [arXiv:hep-th/0501052].

\bibitem{Bern:2005cq}
  Z.~Bern, L.~J.~Dixon and D.~A.~Kosower,
  arXiv:hep-ph/0507005.

\bibitem{Bern:2005hs}
  Z.~Bern, L.~J.~Dixon and D.~A.~Kosower,
  Phys.\ Rev.\ D {\bf 71} (2005) 105013
  [arXiv:hep-th/0501240].


\bibitem{Badger:2005zh}
  S.~D.~Badger, E.~W.~N.~Glover, V.~V.~Khoze and P.~Svrcek,
  JHEP {\bf 0507} (2005) 025
  [arXiv:hep-th/0504159].

\bibitem{Badger:2005jv}
  S.~D.~Badger, E.~W.~N.~Glover and V.~V.~Khoze,
  JHEP {\bf 0601} (2006) 066
  [arXiv:hep-th/0507161].

\bibitem{Bern:1996ja}
  Z.~Bern, L.~J.~Dixon, D.~C.~Dunbar and D.~A.~Kosower,
  Phys.\ Lett.\ B {\bf 394} (1997) 105 
  [arXiv:hep-th/9611127].

\bibitem{Forde:2005ue}
  D.~Forde and D.~A.~Kosower,
  arXiv:hep-th/0507292.

\bibitem{Rodrigo:2005eu}
  G.~Rodrigo,
  JHEP {\bf 0509} (2005) 079
  [arXiv:hep-ph/0508138].

\bibitem{Schwinn:2006ca}
  C.~Schwinn and S.~Weinzierl,
  arXiv:hep-th/0602012.





\end{thebibliography}
\end{document}